\documentclass{emulateapj}

\newcommand{\Dnu}{\mbox{$\Delta \nu$}}

\newcommand{\cms}{\mbox{cm\,s$^{-1}$}}

\newcommand{\ms}{\mbox{m\,s$^{-1}$}}
\newcommand{\kms}{\mbox{km\,s$^{-1}$}}
\newcommand{\muHz}{\mbox{$\mu$Hz}}

\newcommand{\new}[1]{{\bf #1}}
\renewcommand{\new}[1]{\relax #1}

\slugcomment{to appear in ApJ}

\shorttitle{Oscillations in Procyon. I. Observations}
\shortauthors{Arentoft et al.}

\begin{document}

\title{
A multi-site campaign to measure solar-like oscillations in Procyon.\\
I. Observations, Data Reduction and Slow Variations}

\author{
Torben Arentoft,\altaffilmark{1}
Hans~Kjeldsen,\altaffilmark{1}
Timothy~R.~Bedding,\altaffilmark{2}
\\
Micha\"el Bazot,\altaffilmark{1,3}
J{\o}rgen~Christensen-Dalsgaard,\altaffilmark{1}
Thomas~H.~Dall,\altaffilmark{4}
Christoffer Karoff,\altaffilmark{1}
\\ 
Fabien~Carrier,\altaffilmark{5}
Patrick~Eggenberger,\altaffilmark{6}
Danuta Sosnowska,\altaffilmark{7}
\\ 
Robert A. Wittenmyer,\altaffilmark{8}
Michael Endl,\altaffilmark{8}
Travis S. Metcalfe,\altaffilmark{9}
\\ 
Saskia Hekker,\altaffilmark{10,11}
Sabine Reffert,\altaffilmark{12}
\\ 
R.~Paul~Butler,\altaffilmark{13}
Hans~Bruntt,\altaffilmark{2}
L\'aszl\'o~L.~Kiss,\altaffilmark{2}
Simon~J.~O'Toole,\altaffilmark{14}
\\ 
Eiji Kambe,\altaffilmark{15}
Hiroyasu Ando,\altaffilmark{16}
Hideyuki Izumiura,\altaffilmark{15}
Bun'ei Sato,\altaffilmark{17}
\\ 
Michael Hartmann,\altaffilmark{18}
Artie Hatzes,\altaffilmark{18}
\\ 
Francois~Bouchy,\altaffilmark{19}
Benoit~Mosser,\altaffilmark{20}
Thierry~Appourchaux,\altaffilmark{21}
Caroline~Barban,\altaffilmark{20}
Gabrielle~Berthomieu,\altaffilmark{22} 
Rafael~A.~Garcia,\altaffilmark{23}
Eric~Michel,\altaffilmark{20}
Janine~Provost,\altaffilmark{22} 
Sylvaine~Turck-Chi\`eze,\altaffilmark{23}
\\ 
Milena~Marti{\'c},\altaffilmark{24}
Jean-Claude~Lebrun,\altaffilmark{24}
Jerome~Schmitt,\altaffilmark{25}
Jean-Loup~Bertaux,\altaffilmark{24}
\\ 
Alfio~Bonanno,\altaffilmark{26}
Serena~Benatti,\altaffilmark{27}
Riccardo~U.~Claudi,\altaffilmark{27}
Rosario~Cosentino,\altaffilmark{26}
Silvio~Leccia,\altaffilmark{28}
\\ 
S{\o}ren~Frandsen,\altaffilmark{1}
Karsten~Brogaard,\altaffilmark{1}
Lars Glowienka,\altaffilmark{1}
Frank~Grundahl\altaffilmark{1} and
Eric~Stempels\altaffilmark{29}
}

\altaffiltext{1}
{Danish AsteroSeismology Centre (DASC), Department of Physics and
Astronomy, University of Aarhus, DK-8000 Aarhus C, Denmark;
toar@phys.au.dk, hans@phys.au.dk, jcd@phys.au.dk, karoff@phys.au.dk, srf@phys.au.dk,
kfb@phys.au.dk, f002769@phys.au.dk, fgj@phys.au.dk}

\altaffiltext{2}
{Institute of Astronomy, School of Physics, University of Sydney, NSW 2006, Australia;
bedding@physics.usyd.edu.au, bruntt@physics.usyd.edu.au,
laszlo@physics.usyd.edu.au}

\altaffiltext{3}
{Centro de Astrof{\'\i}sica da
Universidade do Porto, Rua das Estrelas, 4150-762 Porto, Portugal;
bazot@astro.up.pt}

\altaffiltext{4}
{Gemini Observatory, 670 N. A'ohoku Pl., Hilo, HI 96720, USA;
tdall@gemini.edu}

\altaffiltext{5}
{Instituut voor Sterrenkunde, Katholieke Universiteit Leuven,
Celestijnenlaan 200 B, 3001 Leuven, Belgium; fabien@ster.kuleuven.be}

\altaffiltext{6}
{Observatoire de Gen\`eve, Ch.~des Maillettes 51, CH-1290 Sauverny,
Switzerland; patrick.eggenberger@obs.unige.ch}

\altaffiltext{7}
{Laboratoire d'astrophysique, EPFL Observatoire CH-1290 Versoix;
danuta.sosnowska@epfl.ch}

\altaffiltext{8}
{McDonald Observatory, University of Texas at Austin, Austin, TX 78712,
USA; robw@astro.as.utexas.edu, mike@astro.as.utexas.edu}

\altaffiltext{9}
{High Altitude Observatory, National Centre for Atmospheric Research,
Boulder, CO 80307-3000 USA; travis@ucar.edu}

\altaffiltext{10} 
{Leiden Observatory, Leiden University, 2300 RA Leiden, The Netherlands}

\altaffiltext{11} 
{Royal Observatory of Belgium, 1180 Brussels, Belgium; saskia@oma.be}

\altaffiltext{12}
{ZAH-Landessternwarte, 69117 Heidelberg, Germany;
sreffert@lsw.uni-heidelberg.de}

\altaffiltext{13}
{Carnegie Institution of Washington, Department of Terrestrial Magnetism,
5241 Broad Branch Road NW, Washington, DC 20015-1305; paul@dtm.ciw.edu}

\altaffiltext{14}
{Anglo-Australian Observatory, P.O.\,Box 296, Epping, NSW 1710, Australia;
otoole@aaoepp.aao.gov.au}

\altaffiltext{15}
{Okayama Astrophysical Observatory, National Astronomical Observatory of
Japan, National Institutes of Natural Sciences, 3037-5 Honjyo, Kamogata,
Asakuchi, Okayama 719-0232, Japan; kambe@oao.nao.ac.jp,
izumiura@oao.nao.ac.jp}

\altaffiltext{16}
{National Astronomical Observatory of Japan, National Institutes of Natural
Sciences, 2-21-1 Osawa, Mitaka, Tokyo 181-8588, Japan;
ando@optik.mtk.nao.ac.jp}

\altaffiltext{17}
{Global Edge Institute, Tokyo Institute of Technology
 2-12-1-S6-6, Ookayama, Meguro-ku, Tokyo 152-8550, Japan;
 sato.b.aa@m.titech.ac.jp}

\altaffiltext{18} 
{Th\"uringer Landessternwarte Tautenburg, Sternwarte 5, 07778 Tautenburg,
Germany; michael@tls-tautenburg.de, artie@tls-tautenburg.de}

\altaffiltext{19}
{Institut d'Astrophysique de Paris, UMR7095, Universit{\'e} Pierre \& Marie
Curie, 98$^{bis}$ Bd Arago, 75014 Paris, France;
bouchy@iap.fr}

\altaffiltext{20}
{Observatoire de Paris, LESIA, UMR 8109, F-92195 Meudon, France;
benoit.mosser@obspm.fr, caroline.barban@obspm.fr, eric.michel@obspm.fr}

\altaffiltext{21}
{Institut d'Astrophysique Spatiale, Universit{\'e} Paris XI-CNRS,
  B{\^a}timent 121, 91405 Orsay cedex, France; Thierry.Appourchaux@ias.u-psud.fr} 

\altaffiltext{22}
{Laboratoire Cassiop{\'e}e, UMR CNRS 6202, Observatoire de la C{\^o}te
  d'Azur, BP 4229, 06304 Nice cedex 4, France;
  Gabrielle.Berthomieu@obs-nice.fr, Janine.Provost@obs-nice.fr}

\altaffiltext{23} 
{DAPNIA/DSM/Service d'Astrophysique, CEA/Saclay, 91191 Gif-sur-Yvette
Cedex, France; rafael.garcia@cea.fr, sylvaine.turck-chieze@cea.fr}

\altaffiltext{24}
{Service d'A{\'e}ronomie du CNRS, BP 3, 91371 Verri{\`e}res le Buisson,
France; milena.martic@aerov.jussieu.fr,
jean-claude.lebrun@aerov.jussieu.fr,  jean-loup.bertaux@aerov.jussieu.fr}

\altaffiltext{25}
{Observatoire de Haute Provence, 04870 St Michel l'Observatoire, France;
jerome.schmitt@oamp.fr}

\altaffiltext{26}
{INAF - Osservatorio Astrofisico di Catania, via S. Sofia 78, 95123
Catania, Italy; abo@oact.inaf.it, rco@ct.astro.it}

\altaffiltext{27}
{INAF - Astronomical Observatory of Padua, Vicolo Osservatorio 5, 35122
Padova, Italy; serena.benatti@oapd.inaf.it, riccardo.claudi@oapd.inaf.it}

\altaffiltext{28}
{INAF - Astronomical Observatory of Capodimonte, Salita Moiariello 16,
80131 Napoli, Italy; leccia@na.astro.it}

\altaffiltext{29}
{School of Physics \& Astronomy, University of St.\ Andrews, North Haugh,
St.\ Andrews KY16 9SS, Scotland; Eric.Stempels@st-andrews.ac.uk}

\begin{abstract} 
We have carried out a multi-site campaign to measure
oscillations in the F5 star Procyon~A.  We obtained high-precision velocity
observations over more than three weeks with eleven telescopes, with almost
continuous coverage for the central ten days.  This represents the most
extensive campaign so far organized on any solar-type oscillator.  We
describe in detail the methods we used for processing and combining the
data.  These involved calculating weights for the velocity time series from
the measurement uncertainties and adjusting them in order to minimize the
noise level of the combined data.  The time series of velocities for
Procyon shows the clear signature of oscillations, with a plateau of excess
power that is centred at 0.9\,mHz and is broader than has been seen for
other stars.  The mean amplitude of the radial modes is
$38.1\pm1.3$\,\cms{} (2.0 times solar), which is consistent with previous
detections from the ground and by the WIRE spacecraft, and also with the
upper limit set by the MOST spacecraft.  The variation of the amplitude
during the observing campaign allows us to estimate the mode lifetime to be
$1.5_{-0.8}^{+1.9}$\,d.  We also find a slow variation in the radial
velocity of Procyon, with good agreement between different telescopes.
These variations are remarkably similar to those seen in the Sun, and we
interpret them as being due to rotational modulation from active regions on the
stellar surface.  The variations appear to have a period of about 10 days,
which presumably equals the stellar rotation period or, perhaps, half of
it.  The amount of power in these slow variations indicates that the
fractional area of Procyon covered by active regions is slightly higher than for
the Sun.
\end{abstract}


\section{Introduction}

Measuring solar-like oscillations in main-sequence and subgiant stars
requires high-precision observations -- either with spectroscopy or
photometry -- combined with coverage that is as continuous as possible.
Most of the results have come from high-precision Doppler measurements
using ground-based spectrographs, while measurements from spacecraft have
also been reported (see \citealt{B+K2007c} and \citealt{AChDC2008} for
recent summaries).

Procyon has long been a favourite target for oscillation searches.  At
least eight separate velocity studies have reported an excess in the power
spectrum, beginning with that by \citet{BGN91}, which was the first report
of a solar-like power excess in another star.  For the most recent
examples, see \citet{MLA2004}, \citet{ECB2004}, \citet{BMM2004} and
\citet{LKB2007}.  These studies agreed on the location of the excess power
(around 0.5--1.5\,mHz) but they disagreed on the individual oscillation
frequencies.  However, a consensus has emerged that the large separation
(the frequency separation between consecutive overtone modes of a given
angular degree) is about 55\,\muHz. Evidence for this value was first
given by \citet{MMM98} and the first clear detection was made by
\citet{MSL99}.

Controversy was generated when photometric observations obtained with the
MOST satellite failed to reveal evidence for oscillations
\citep{MKG2004,GKR2007,BAB2008}.  However, \citet{BKB2005} argued that the
MOST non-detection was consistent with the ground-based data.  Meanwhile,
\citet{R+RC2005} suggested that the signature of oscillations is indeed
present in the MOST data at a low level (see also \citealt{Mar08}).  Using
space-based photometry with the WIRE satellite, \citet{BKB2005b} extracted
parameters for the stellar granulation and found evidence for an excess due
to oscillations.

All published velocity observations of Procyon have been made from a single
site, with the exception of two-site observations by \citet{MLA2004}.  Here
we describe a multi-site campaign on Procyon carried out in 2007 January,
which was the most extensive velocity campaign so far organized on any
solar-type oscillator.  The only other comparable effort to measure
oscillations in this type of star was the multi-site photometric campaign
of the open cluster M67 \citep{GBK93}.

\section{Velocity Observations}

We observed Procyon from 2006 December 28 until 2007 January 23, using a
total of eleven telescopes at eight observatories.  These are listed in
Table~\ref{tab.sites}, ordered westward by longitude.  Note that the FIES
spectrograph on the Nordic Optical Telescope was still being commissioned
during the observations and the velocity precision is therefore somewhat
lower than for the other telescopes.

The team members from each telescope were responsible for producing a
velocity time series from the observations, together with estimates of
uncertainties.  In six of the spectrographs, the stellar light was passed
through an iodine absorption cell to provide a stable wavelength reference.
In four others, wavelength calibration was achieved by recording the
spectrum from a thorium-argon emission lamp alongside the stellar spectrum,
while with EMILIE, exposures of the stellar spectrum were alternated with
those of a white-light source passing through an iodine cell.  Details of
the methods used with each spectrograph are given in the references listed
in Table~\ref{tab.sites} and details of the observations are given in
Table~\ref{tab.obs}.

\begin{figure*}
\epsscale{0.8}
\plotone{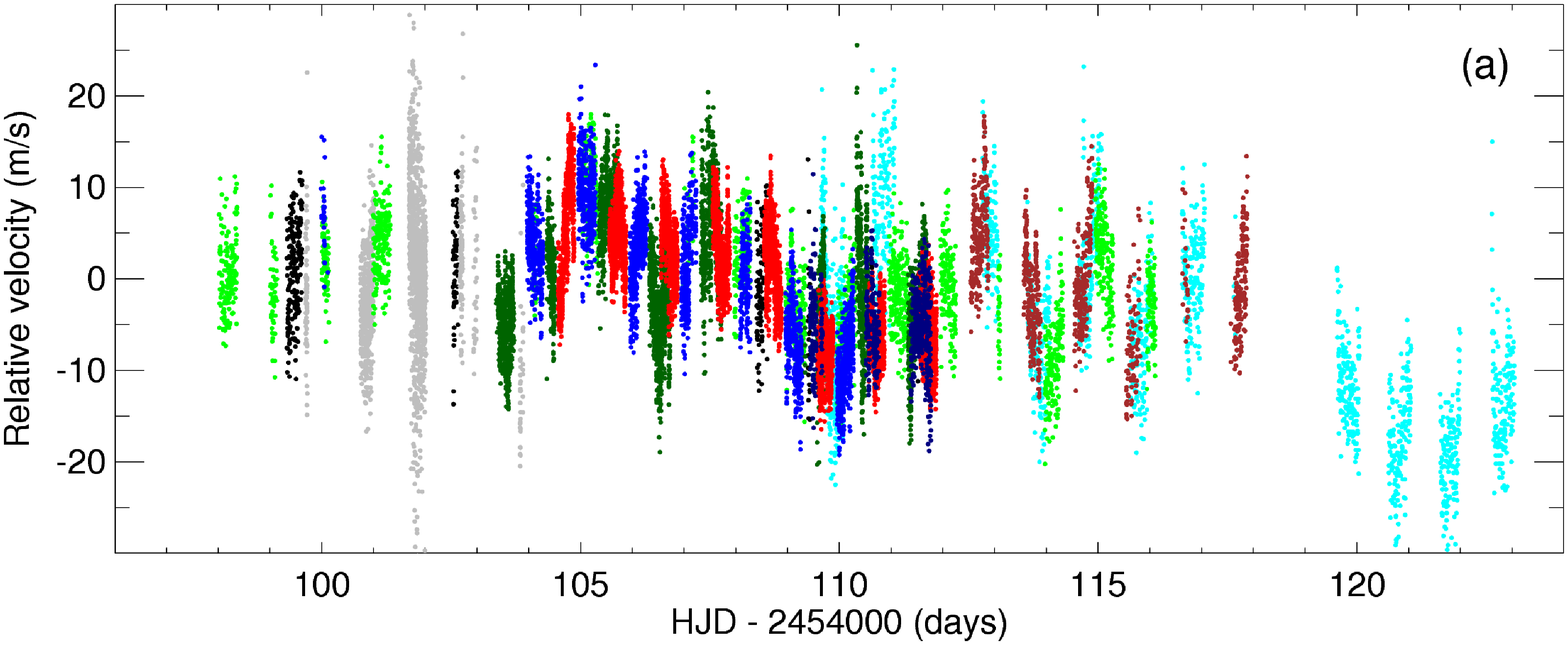}
\plotone{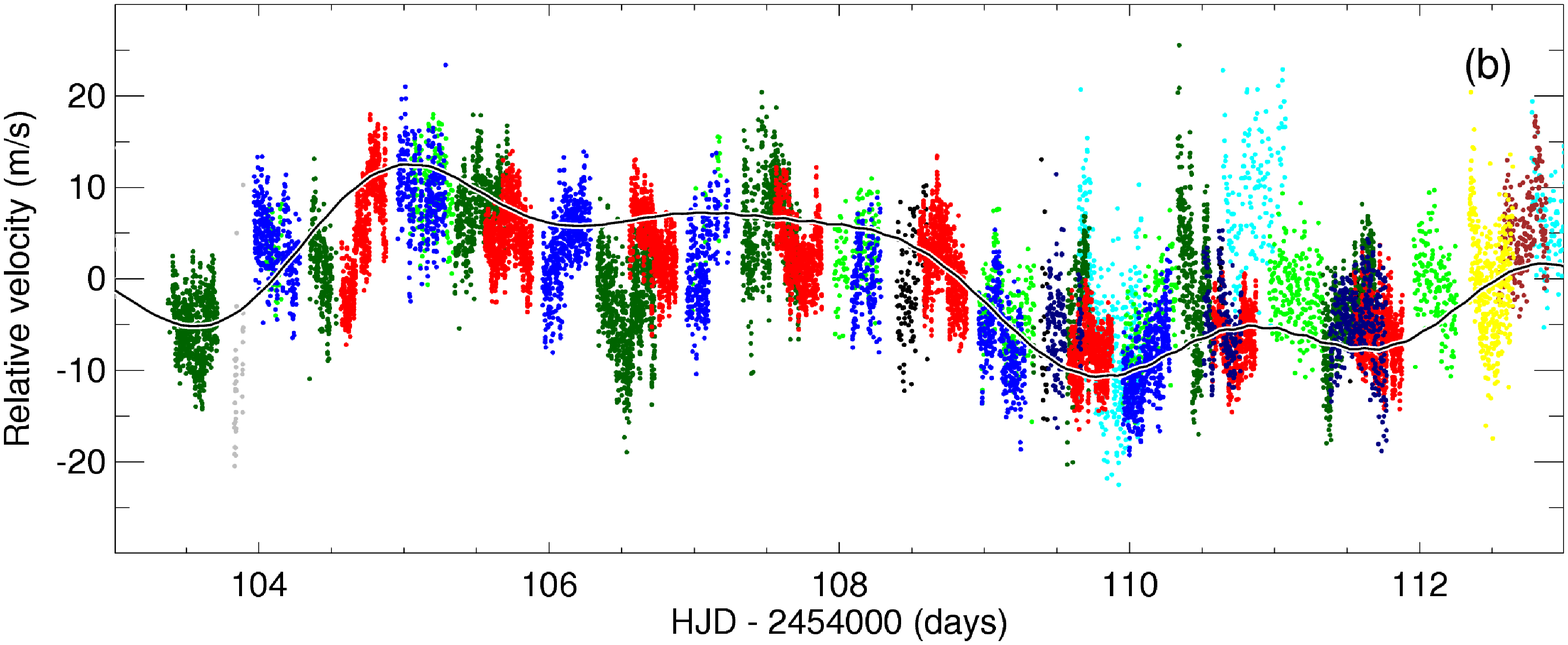}
\plotone{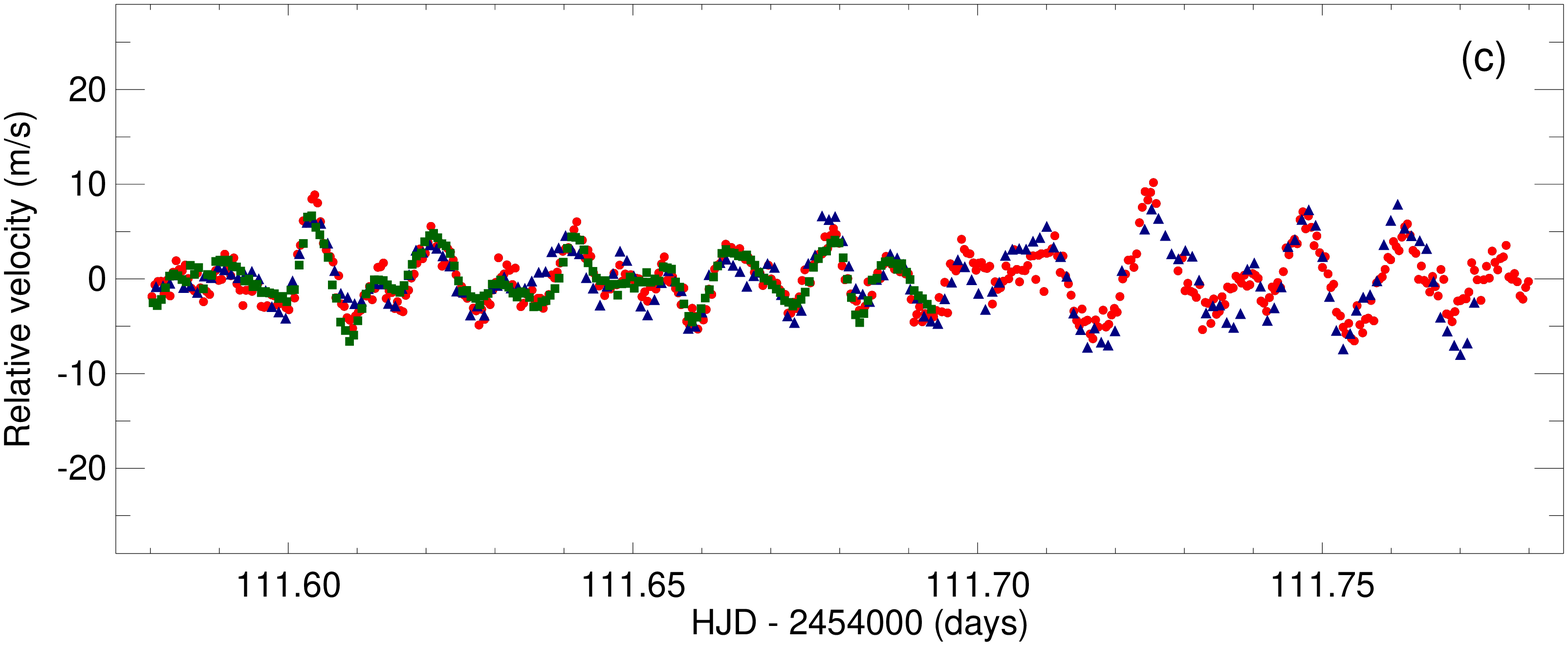}
\plotone{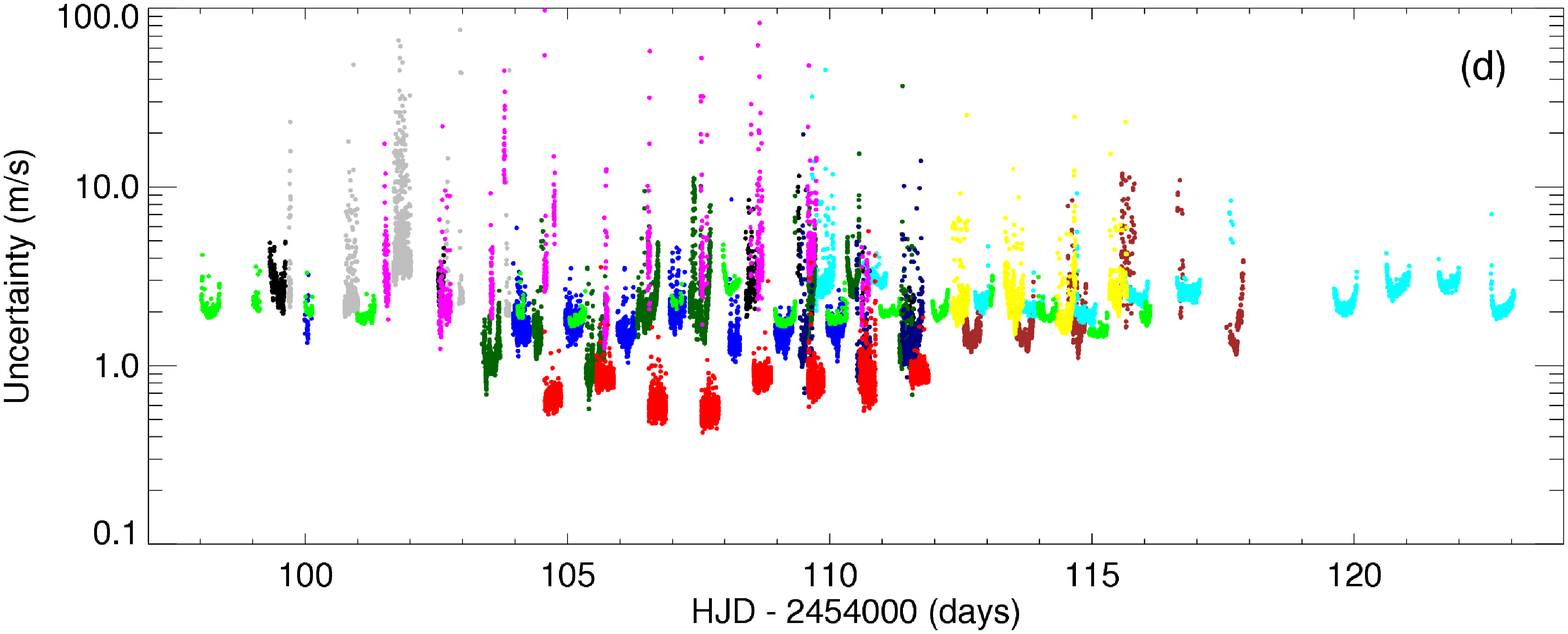}
\caption[]{\label{fig.series} Velocity measurements of Procyon, color-coded
as follows: 
HARPS         = red;
CORALIE       = brown;
McDonald      = gray;
Lick          = cyan;
UCLES         = blue;
Okayama       = green;
Tautenburg    = black;
SOPHIE        = dark green;
SARG          = dark blue;
FIES          = magenta;
EMILIE        = yellow.

({\em a})~The full time series, before any removal of slow trends (EMILIE
and FIES are not shown).

({\em b})~Close-up of the central ten days (FIES not shown).  

({\em c})~Close-up of a five-hour segment during which three spectrographs
observed simultaneously: HARPS (red circles), SOPHIE (dark green squares) and
SARG (dark blue triangles).  All three series have been high-pass filtered
to remove slow trends and the SOPHIE and SARG data have been smoothed
slightly (using a boxcar with a width of three data points).

({\em d})~The time series of the final noise-optimized uncertainties,
showing all 11 telescopes.
}
\end{figure*}

The velocity time series of Procyon is shown in Fig.~\ref{fig.series}{\em
a\/}, using a different color for each telescope.  Differences between
telescopes in the absolute zero point of velocity are not significant, and
so all the curves have been shifted into alignment by subtracting a
constant offset.  This was done by setting the velocities from each
telescope to have zero mean, excepting Lick and SARG, for which better
alignment was achieved by setting the means to $-6$\,\ms{} and $-5$\,\ms,
respectively.  Note that EMILIE is not shown in Fig.~\ref{fig.series}{\em
a\/} because those data were referenced to a different value on each night
and so the night-to-night variations are not measurable (this does not
affect their usefulness for oscillations studies, however).  FIES is also
not shown in Fig.~\ref{fig.series}{\em a\/} because it has much greater
scatter than the rest.

In Fig.~\ref{fig.series}{\em a\/} we see variations in the radial velocity
of Procyon on timescales of days.  The good agreement between the different
telescopes indicates that these slow variations have a stellar origin,
although the imperfect match in overlapping sections shows that there is
also a contribution from instrumental drifts.  Figure~\ref{fig.series}{\em
b\/} shows a close-up of the central part of the campaign, during which the
coverage was above 90\%.  The solid curve shows the velocities after
smoothing, to better reveal the slow variations, which are discussed
in~\S\ref{sec.slow-var} below.

While interesting in their own right, the slow variations in the velocity
series significantly affect our ability to detect oscillations, due to
spectral leakage of power from the low-frequency part of the spectrum to
the oscillation region at higher frequencies.  Before merging the data from
the individual telescopes, we therefore removed the low-frequency
variations from the velocity series.  This was done for each telescope by
removing all the power below 280$\mu$Hz, a value that was chosen so as to
effectively remove the slow variations without affecting the oscillation
signal.  This filtering was done by subtracting a smoothed version of the
time series that contained all the power below that cut-off frequency.

In Fig.\ref{fig.series}{\em c\/} we show a close-up of a segment during
which three spectrographs were observing simultaneously.  The stellar
oscillations are clearly visible, with typical periods of about 15 minutes,
and there is good agreement between the different telescopes.  Note that
these data have been filtered to remove the slow variations.

\section{Optimizing the Weights}     \label{sec.weights}

The procedures for extracting velocities for each telescope also produced
estimates of the uncertainties,~$\sigma_i$.  In our analysis, we used these
uncertainties to calculate noise-optimized weights in the usual way, namely
$w_i=1/\sigma_i^2$.  If weights are not used when calculating the power
spectrum, a few bad data points can dominate and increase the noise floor
significantly.

We now describe the process we used to adjust these weights, which aims to
minimize the noise level in the final power spectrum.  The procedure
involves identifying and revising those uncertainties that were too
optimistic, and at the same time rescaling the uncertainties to be in
agreement with the actual noise levels in the data.  These methods have
already been described in previous papers \citep{BBK2004,BKA2007,LKB2007},
but the present analysis differs slightly from those descriptions and we
will therefore describe them in some detail.  One difference is that the
analysis had to be tailored to the individual time series because of the
large range of Nyquist frequencies (see Table~\ref{tab.obs}).

\begin{figure*}
\epsscale{0.9}
\plotone{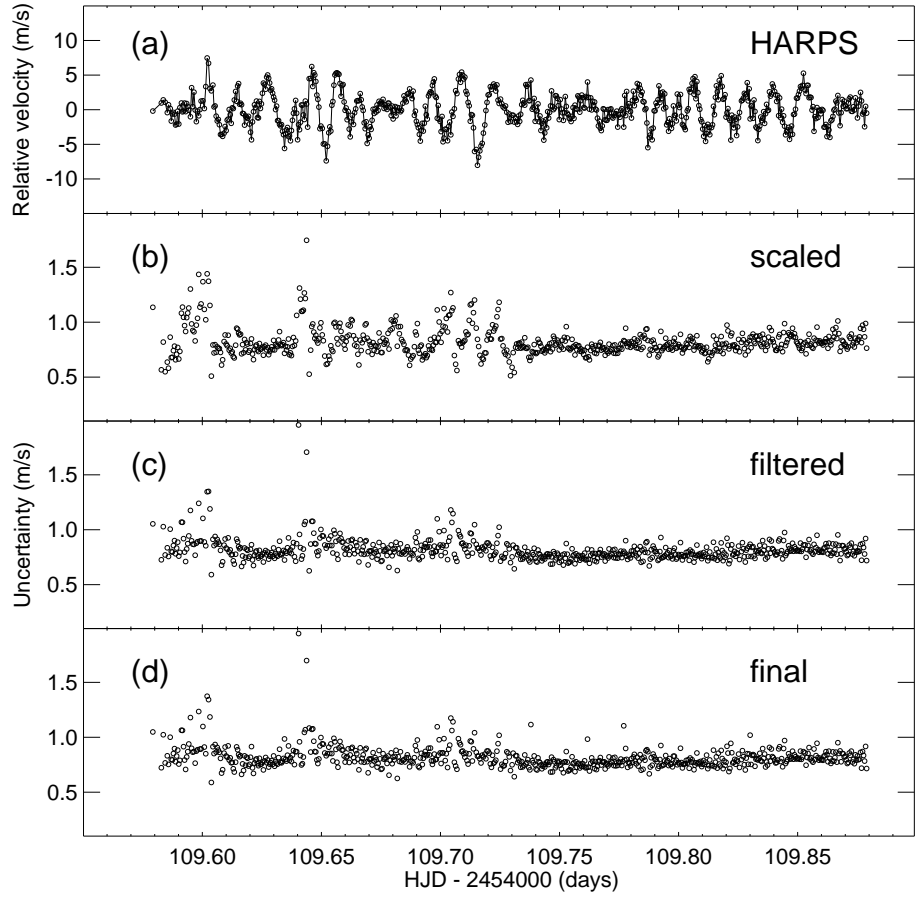}
\caption[]{\label{fig.show.harps} Steps in the adjustments of weights,
illustrated using HARPS data from a single night. 
({\em a})~The velocities, with the slow variations removed.
({\em b})~The uncertainties, after scaling to satisfy
Eq.~\ref{eq.condition} but before any further adjustments.
({\em c})~The uncertainties after filtering to remove power on the
timescale of the oscillations.
({\em d})~The final uncertainties, after adjusting to optimize the noise
(see text).
}
\end{figure*}

\begin{figure*}
\epsscale{0.9}
\plotone{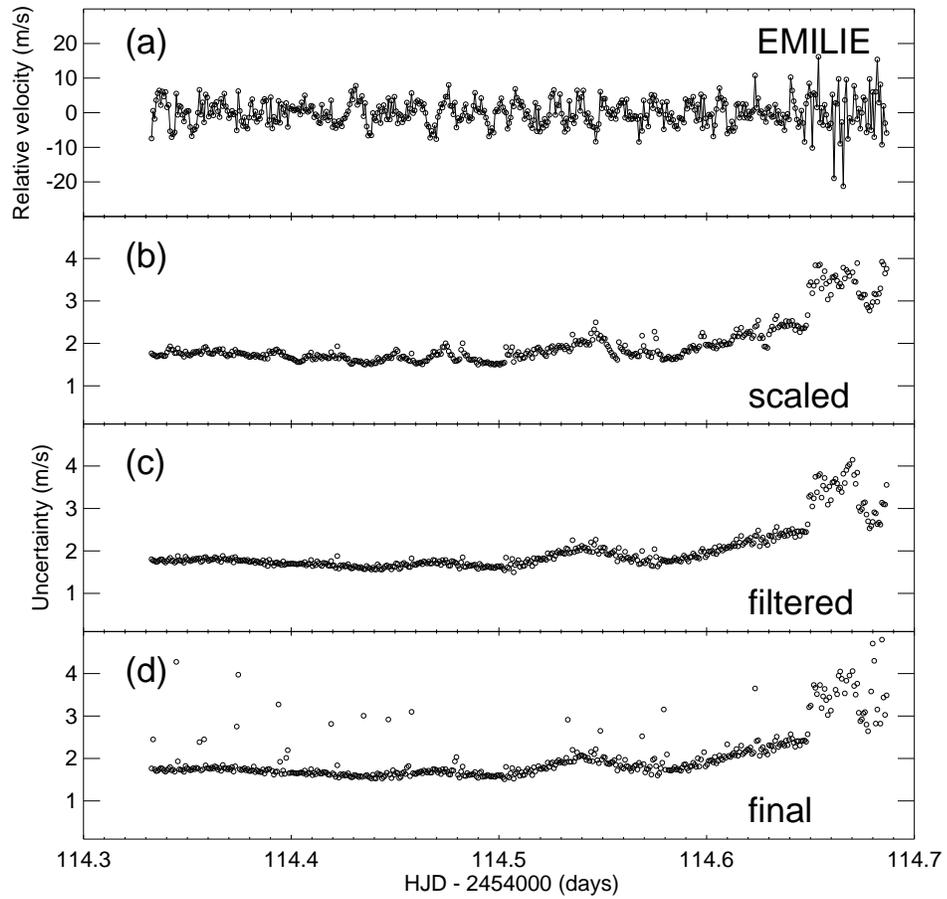}
\caption[]{\label{fig.show.emilie} Same as Fig.~\ref{fig.show.harps}, but
  for a single night from EMILIE.  
}
\end{figure*}

To illustrate the process, we show in Figs.~\ref{fig.show.harps}
and~\ref{fig.show.emilie} segments of data at different stages in the
process for two telescopes (HARPS and EMILIE).  The top panels
(Figs.~\ref{fig.show.harps}{\em a\/} and~\ref{fig.show.emilie}{\em a\/})
show the velocities for a single night, with the slow variations removed.
The remaining panels show the uncertainties at different stages in the
analysis.

It is important to stress that we are not adjusting the velocities, only
the uncertainties.  Of course, those adjustments still affect the power
spectrum of the velocities (which is, after all, why we are making the
adjustments) and so it is important to ensure that they do not distort the
oscillation signal and that the final weights reflect as accurately as
possible the actual noise properties of the series.

\subsection{Scaling the uncertainties}

We have scaled the uncertainties so that they agree with the noise level in
the corresponding amplitude spectrum, $\sigma_{\rm amp}$, as measured at
high frequencies.  This was done for each night and each telescope by
multiplying the uncertainties, $\sigma_i$, by a constant so that they
satisfied equation~(3) of \citet{BBK2004}:
\begin{equation}
      \sigma_{\rm amp}^2 \sum_{i=1}^{N} \sigma_i^{-2}  = \pi. 
        \label{eq.condition}
\end{equation}
This scaling was repeated after each step in the process described below.
Figures~\ref{fig.show.harps}{\em b\/} and~\ref{fig.show.emilie}{\em b\/}
show the uncertainties after scaling, and before any further adjustments.

\subsection{Filtering the uncertainties}     \label{sec.init-chk}

It is clear that the uncertainties in some parts of the time series show
variations that correlate with the oscillations of Procyon.  The clearest
example is HARPS, as shown by comparing the top two panels of
Fig.~\ref{fig.show.harps}, but the effect is also visible for other
telescopes (e.g., Fig~\ref{fig.show.emilie}).  To remove this structure in
the uncertainties, we have bandpass-filtered each of the uncertainty time
series to remove all power in the frequency range 280--2200$\mu$Hz.  This
removed fluctuations in the weights on the timescale of the stellar
oscillations, while retaining information on longer timescales (such as
poorer conditions at the beginnings and ends of nights) and on shorter
timescales (such as individual bad data points).  This process resulted in
slightly lower noise levels in the final power spectra for some of the
individual telescopes, reflecting the fact that the uncertainties, like any
measurement, contain noise that is reduced by bandpass filtering.
Figures~\ref{fig.show.harps}{\em c\/} and~\ref{fig.show.emilie}{\em c\/}
show the uncertainties after filtering.

We also noticed a few data points ($\sim$10) with unrealistically {\em
low\/} uncertainties.  These points would be given too high a weight in the
analysis and the uncertainties were therefore reset to the mean uncertainty
for that telescope night.

\subsection{Down-weighting bad data points}

Down-weighting of bad data points was done following the method described
by \citet{BBK2004}, with one difference that is discussed below.  The first
step was to make a high-pass-filtered version of the velocity time series
in which both the slow variations and the stellar oscillations were
removed.  This gave us a series of residual velocities, $r_i$, in which we
could identify data points that needed to be down-weighted, without being
affected by spectral leakage from the oscillations.  The frequency limit of
this high-pass filter varied from telescope to telescope, depending on the
Nyquist frequency of the data.

We compared the velocity residuals, $r_i$, with the corresponding
uncertainty estimates, $\sigma_i$.  Bad data points are those for which the
ratio $|r_i/\sigma_i|$ is large, i.e., where the residual velocity deviates
from zero by more than expected from the uncertainty estimate.
\citet{BBK2004}, who analyzed data similar to ours, found that the fraction
of good data points was essentially unity up to $|r_i/\sigma_i|=2$ and then
dropped off quickly for larger values of $|r_i/\sigma_i|$ -- see Figure~3
of \citet{BBK2004}.  They therefore introduced the factor $f$, which is the
fraction of good data points as a function of $|r_i/\sigma_i|$ and which
they obtained as the ratio between the distribution of data points in a
cumulative histogram of $|r_i/\sigma_i|$ and a best-fit Gaussian
distribution.


We used a slightly different approach.  With the knowledge that points with
large values of $|r_i/\sigma_i|$ are bad, we introduced an analytical
function
\begin{equation}
  f(x_i)=\frac{1}{1+\left(\displaystyle{x_i\over x_0} \right)^{10}},\quad x_i=|r_i/\sigma_i|,
  \label{eq.fx}
\end{equation}
which has shape very similar to the fraction $f$ as a function of
$|r_i/\sigma_i|$ in \citet{BBK2004}.  The adjustable parameter $x_0$
controls the amount of down-weighting; it sets the value of
$|r_i/\sigma_i|$ for which the weights are multiplied by 0.5, and so it
determines how bad a data point should be before it is down-weighted.  The
optimum choice for $x_0$ was found through iteration, as described below.
Once this was done, we used $f(x_i)$ to adjust the weights by dividing
$\sigma_i$ by $\sqrt{f(x_i)}$, as in \citet{BBK2004}.



The noise level used for optimizing $x_0$ was measured in a frequency band
near 2\,mHz in a weighted amplitude spectrum, between the oscillations and
the high-frequency part of the spectrum used for determining the
$|r_i/\sigma_i|$ values.  The exact position of the band was chosen for
each spectrograph separately, because of the differences in Nyquist
frequencies.  For each trial value of $x_0$, the noise level was determined
from a time series in which all power had been removed at both the low- and
high-frequency side of the frequency band, i.e., from a bandpass-filtered
time series containing information in the specific frequency band only.
This was done because applying the weights changes the spectral window
function and thus the amount of spectral leakage into the frequency band
where we determine the noise: if we did not filter out the oscillations and
the high-frequency part of the spectrum, we would optimize for a
combination of low noise {\em and\/} minimum amount of spectral leakage
(from both the low- and high frequency side of the passband).  In other
words, the spectral window function would influence our choice of $x_0$,
which is not optimal for obtaining the lowest possible noise level.

The procedure described above was repeated for a range of $x_0$ values and
we chose the one that resulted in the lowest noise in the power spectrum.
Depending on the telescope, and hence the noise properties of the time
series, the optimal values of $x_0$ ranged from 1.7 to 4.3.
Figures~\ref{fig.show.harps}{\em d\/} and~\ref{fig.show.emilie}{\em d\/}
show the final uncertainties for HARPS and EMILIE.  

This completes our description of the process used to adjust the
uncertainties.  The results from calculating weighted power spectra using
these uncertainties are presented in~\S\ref{sec.power.spectra}.  First,
however, we discuss the slow variations in the velocity of Procyon that are
present in our data.

\section{Results} 

\subsection{Slow variations in stellar velocity}     \label{sec.slow-var}

\begin{figure*}
\epsscale{0.9}
\plotone{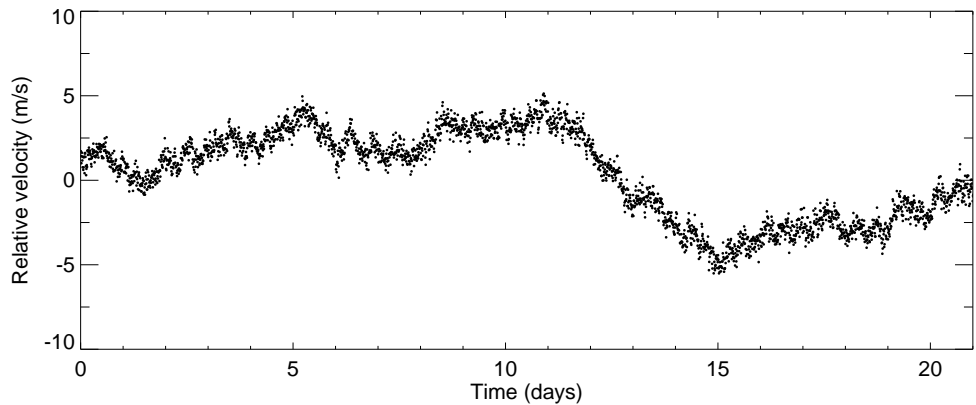}
\caption[]{\label{fig.golf} Time series of velocity measurements of the Sun
obtained over 21 days with the GOLF instrument on the SOHO spacecraft.  }
\end{figure*}

The slow variations in the radial velocity of Procyon seen in
Fig.~\ref{fig.series} are remarkably similar to those seen in the Sun.
This can be seen from Fig.~\ref{fig.golf}, which shows a typical time
series of solar velocity measurements made with the GOLF instrument on the
SOHO spacecraft \citep{UGR2000,GTB2005}.  Longer series of GOLF data show
variations with a period of about 13\,d arising from active regions
crossing the solar disk (Fig.~11 in \citealt{GTB2005}; see also
\citealt{TGK2008}).  \new{This 13-day periodicity in solar velocities was
first observed by \citet{CIMc82}, who attributed it to rapid rotation of
the core, but it was subsequently shown to be due to surface rotation of
active regions \citep{Du+S83,A+M83,E+G83}.}  Similarly, we attribute the
slow variations in the radial velocity of Procyon to the appearance and
disappearance of active regions and their rotation across the stellar disk.
This explanation was also invoked by \citet{MBC2005} to explain much larger
velocity variations measured with HARPS for the COROT target HR~2530 (=
HD~49933; spectral type F5\,V).

The slow variations in Procyon appear to have a period of $P_{\rm slow} =
10.3\pm0.5$\,d, as measured from the highest peak in the power spectrum.
\new{This agrees with an apparent periodicity of about 10\,d in two-site
observations of Procyon obtained over 20 nights by \citet[see their
Fig.~12]{KAS2008}.}  Identifying this as the stellar rotation period and
using a radius of 2.05\,$R_\sun$ \citep{KTM2004} implies a surface
rotational speed at the equator of $v = 10 \pm 0.5$\,\kms.  The measured
value of $P_{\rm slow}$ period might also correspond to half the rotation
period \new{\citep{Cla2003}}, in which case the rotational speed would be
half the value given above.  Combining with the spectroscopic value of $v
\sin i = 3.16 \pm 0.50$\,\kms{} \citep{APAL2002} gives an inclination angle
of $i=18 \pm 3^{\circ}$ (if $P_{\rm rot} = P_{\rm slow}$) and $i=39 \pm
7^{\circ}$ (if $P_{\rm rot} = 2 P_{\rm slow}$).  The inclination of the
binary orbit is $31.1 \pm 0.6^{\circ}$ \citep{GWL2000} and so, if we
require that the rotation axis of Procyon is aligned with the orbital
rotation axis, it may be that $P_{\rm rot} = 2 P_{\rm slow}$.

\subsection{Power spectra and stellar activity} \label{sec.power.spectra}

\begin{figure*}
\epsscale{1.0}
\plotone{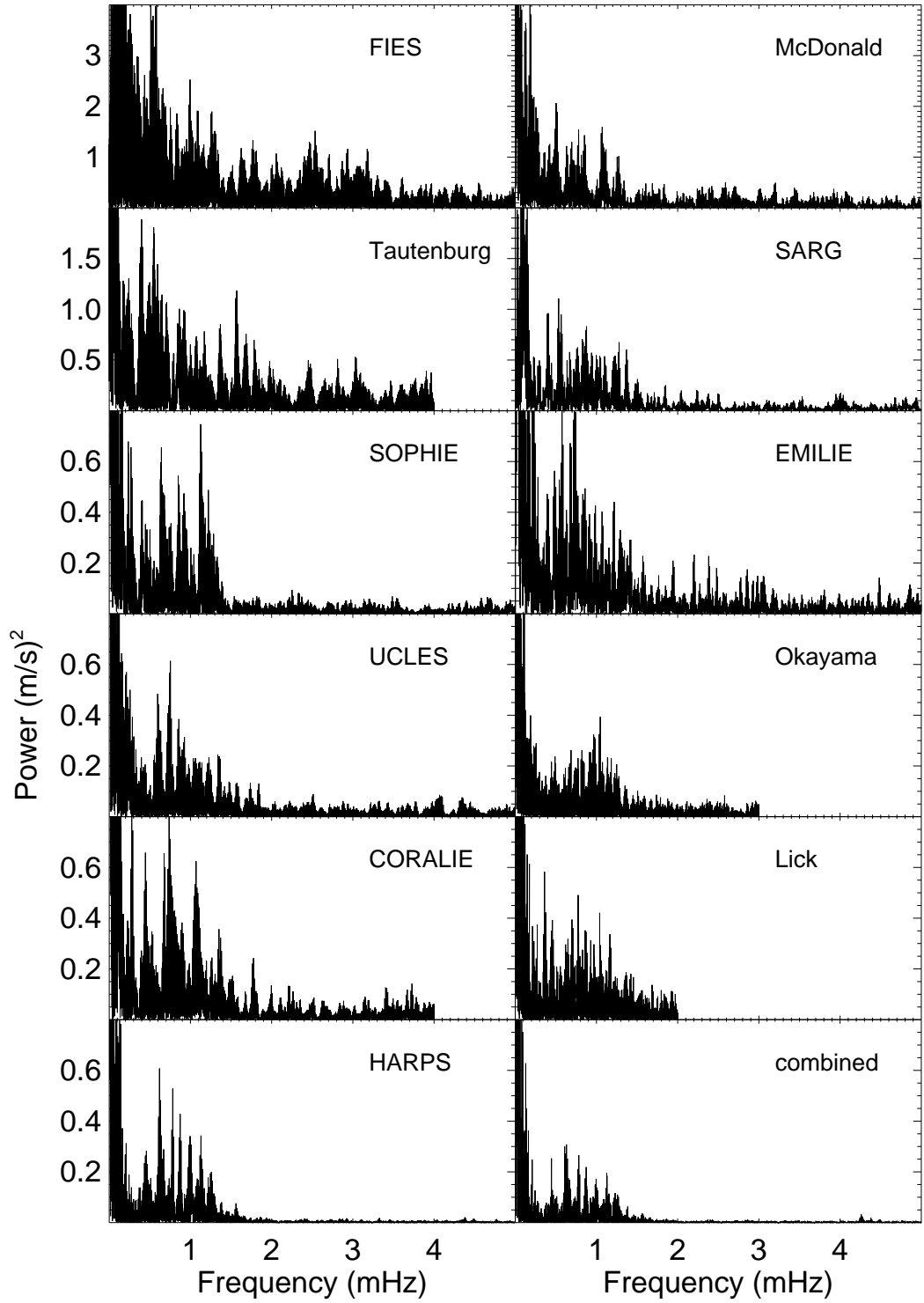}
\caption[]{\label{fig.all.power} Power spectra for all 11 telescopes,
together with that of the combined series.  Note that the vertical scale is
not the same for all panels.}
\end{figure*}

The weighted power spectrum for each telescope and for the combined time
series, based on the uncertainties discussed in~\S\ref{sec.weights}, are
shown in Fig.~\ref{fig.all.power}.  The noise levels, as measured at high
frequencies in the amplitude spectrum ($\sigma_{\rm amp}$), are given in
column~8 of Table~\ref{tab.obs}.  The final column of that table gives the
mean noise level per minute of observing time, with a spread that reflects
a combination of factors, including telescope aperture, observing duty
cycle, spectrograph design, as well as atmospheric conditions such as
seeing.

\begin{figure*}
\epsscale{0.8}
\plotone{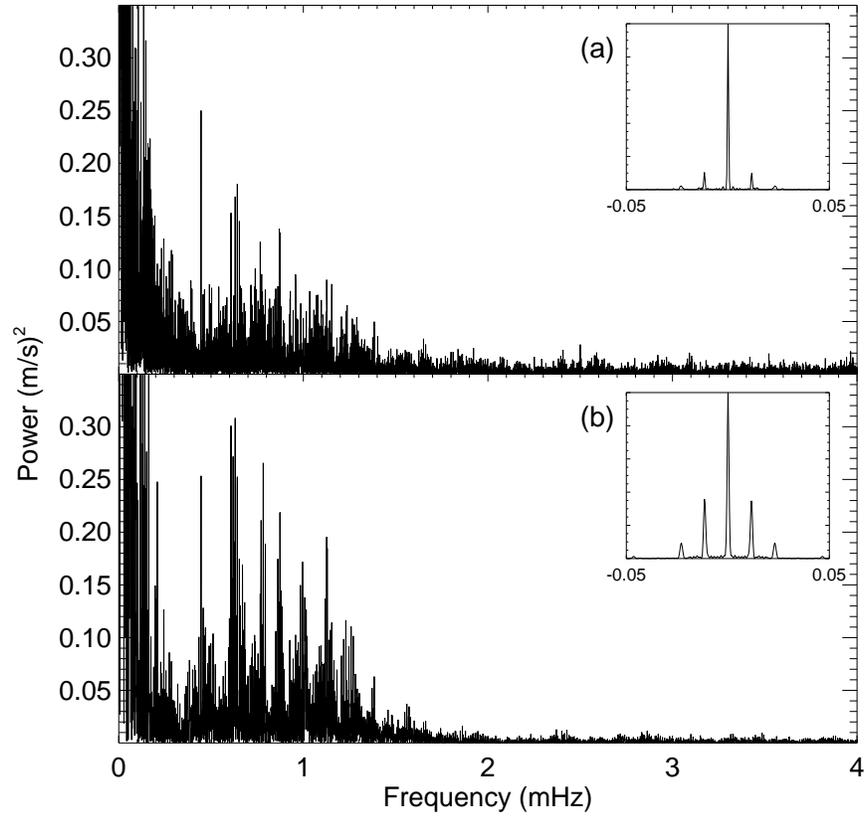}
\caption[]{\label{fig.power} Final power spectrum based on the
noise-optimized weights (lower panel), and also without applying
the weights (upper panel).  The inset shows the spectral window.   }
\end{figure*}

The power spectrum of the combined time series is shown again in
Fig.~\ref{fig.power}, both with and without the use of weights.  As is
well-established, using weights reduces the noise level significantly, at
the cost of increasing the sidelobes in the spectral window (because the
best data segments are given more weight -- see insets).  When weights are
used, the noise level above 3\,mHz is 1.9\,\cms{} in amplitude, but this
does include some degree of spectral leakage from the oscillations.  If we
high-pass filter the spectrum up to 3\,mHz, the noise level drops to
1.5\,\cms{} in amplitude.  Note that without the use of weights, the noise
level is higher by more than a factor of two.

Looking again at Fig.~\ref{fig.power}, we see that the use of weights
appears to have increased the amplitude of the oscillations.  In fact, this
indicates the finite lifetime of the oscillation modes: in
Fig.~\ref{fig.power}{\em b} the HARPS data are given the highest weight,
and so the effective duration of the observations is decreased (and the
sidelobes in the spectral window become much stronger).  Our estimate of
the mode lifetime is given in~\S\ref{sec.lifetime}.

\begin{figure*}
\epsscale{0.8}
\plotone{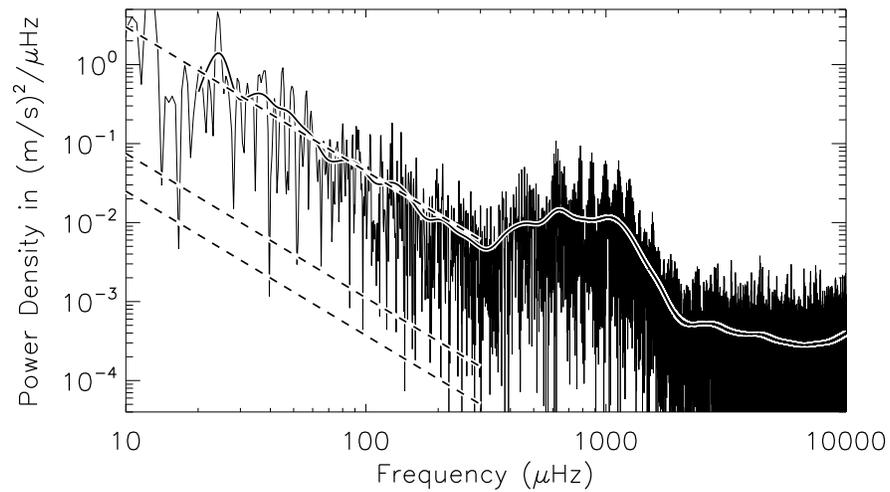}
\caption[]{\label{fig.pds.harps} Power density spectrum of Procyon from the
HARPS data, and the same after smoothing.  The lower two dashed lines show
the solar activity level at minimum and maximum, and the upper line is the
solar maximum activity shifted upwards by a factor of 40.  }
\end{figure*}

\begin{figure*}
\epsscale{0.8}
\plotone{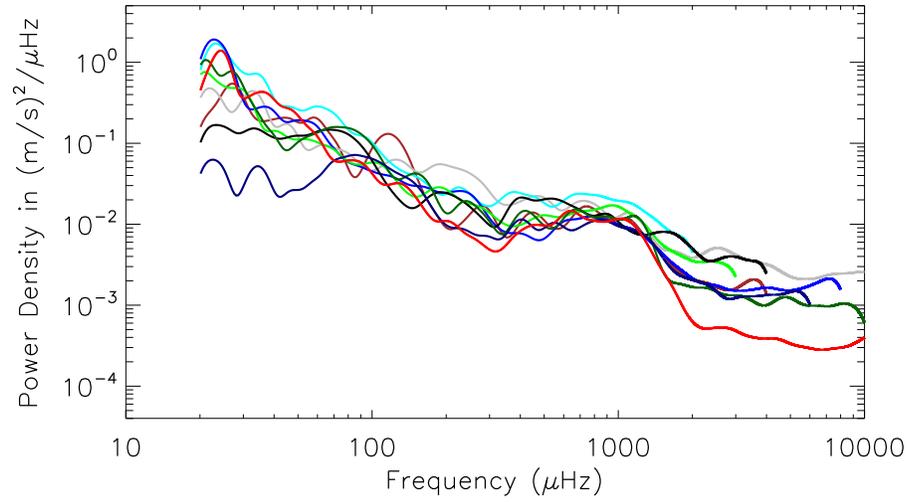}
\caption[]{\label{fig.pds.amp} Smoothed power density spectra (see
Fig~\ref{fig.pds.harps}) for the nine telescopes shown in
Fig.~\ref{fig.series}{\em a\/}, showing a similar background from stellar
noise at low frequencies and different levels of white noise at high
frequencies.  The color coding is the same as in Fig.~\ref{fig.series}. }
\end{figure*}

It is also useful to convert to power density, which is independent of the
observing window and therefore allows us to compare noise levels.  This is
done by multiplying the power spectrum by the effective length of the
observing run, which we calculate as the reciprocal of the area integrated
under the spectral window (in power).  The values for the different
telescopes are given in column~3 of Table~\ref{tab.obs}.  In
Fig.~\ref{fig.pds.harps} we show the power density spectrum on a
logarithmic scale for the HARPS data, which has the lowest noise per minute
of observing time.  We see three components: (i)~the oscillations (about
300--1100\,\muHz); (ii)~white noise at high frequencies; and (iii)~a
sloping background of power at low frequencies (stellar granulation and
activity, and presumably also some instrumental drift).
Figure~\ref{fig.pds.amp} compares the power density spectra for the
different telescopes.  
They show a similar oscillation signal and similar background from stellar
noise at lower frequencies (below about 250\,\muHz), with different levels
of white noise at higher frequencies (above about 2000\,\muHz), reflecting
the different levels of photon noise.

In Fig.~\ref{fig.pds.harps}, the lower two dashed lines indicate the
background level in the Sun, as measured from the GOLF data during solar
minimum and maximum, respectively.  The upper dashed line is the solar
maximum line shifted to match the power density of Procyon, which required
multiplying by a factor of 40.  We can use this scaling factor to estimate
the fraction of Procyon's surface that is covered by active regions,
relative to the Sun, as follows.  The low-frequency part of the velocity
power-density spectrum from the Sun falls off as frequency squared
\citep{Har85,PRCJ99}, and we see the same behaviour for Procyon.  Hence, in
both cases we have
\begin{equation} 
  PD(\nu) \propto \nu^{-2}.
\end{equation} 
Let $T$ be the typical time for an active region to be visible on the
surface (which may depend on both rotation and the typical lifetime of
active regions).  The amplitude of the velocity signal at frequency $\nu_0
= 1/T$ will be proportional to the fractional area covered by active
regions, $da/a$, and to the projected rotational velocity, $v\sin i$.  The
power density at $\nu_0$ is therefore:
\begin{equation} 
  PD(\nu_0) = (da/a)^2 (v\sin i)^2.
\end{equation} 
Combining these gives
\begin{equation} 
  PD(\nu) = \left(\frac{da}{a}\right)^2 \left(\frac{v \sin i}{T}\right)^2 \nu^{-2}.
  \label{eq.spots}
\end{equation} 
We will assume that $T$ is proportional to the rotation period, which we
take to be $10.3$\,d (or twice that value) for Procyon and 25.4\,d for the
Sun.  The measured values for $v \sin i$ are 3.2\,\kms{} for Procyon
\citep{APAL2002} and 2.0\,\kms{} for the Sun.  Combining these values with
our measurement of the power densities indicates that the area covered by
active regions on Procyon is about 1.6 times the solar maximum value (or
twice that value).  No detection of a magnetic field in Procyon has been
reported, and published upper limits imply that the average field cannot be
more than a few times solar (see Table 3 in \citealt{KHV2007}), which
appears to be consistent with our results.

\subsection{Oscillation amplitude and mode lifetime} \label{sec.lifetime}

To measure the amplitude of oscillations in Procyon, we have used the
method described by \citet{KBA2008}.  In brief, this involves the following
steps: (i)~heavily smoothing the power spectrum (by convolving with a
Gaussian having a full width at half maximum of $4\Dnu$, where $\Dnu$
is the large frequency separation), to produce a single hump of excess
power that is insensitive to the fact that the oscillation spectrum has
discrete peaks; (ii)~converting to power density
(see~\S\ref{sec.power.spectra}); (iii)~fitting and subtracting the
background noise; and (iv)~multiplying by $\Dnu/4.09$ and taking the square
root, in order to convert to amplitude per radial oscillation mode.
Note that 4.09 is the effective number of modes per order for
full-disk velocities observations, normalized to the amplitudes of radial
($l=0$) modes -- see \citet{KBA2008} for details.

\begin{figure*}
\epsscale{0.8}
\plotone{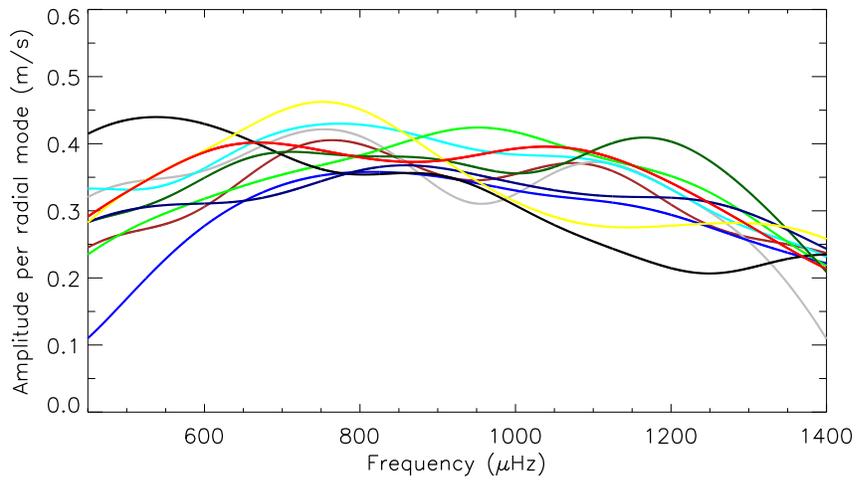}
\caption[]{\label{fig.amp} Smoothed amplitude curves for Procyon for ten
  telescopes, using the same color coding as Fig.~\ref{fig.series}. }
\end{figure*}

\begin{figure*}
\epsscale{0.8}
\plotone{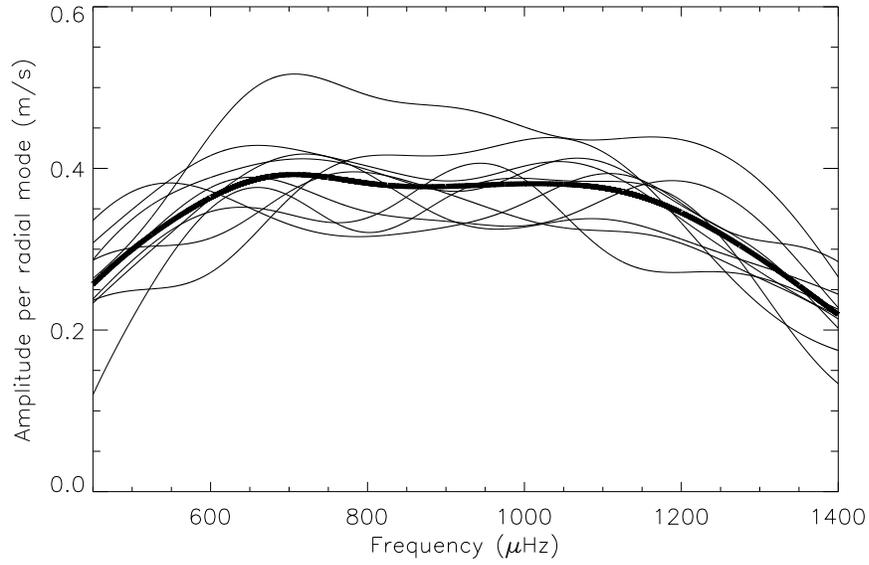}
\caption[]{\label{fig.amp.segments} Smoothed amplitude curves for Procyon
  from ten 2-day segments of the combined time series (thin lines),
  together with their mean (thick line).  }
\end{figure*}

We applied this method to each of the telescopes separately, and the result
is shown in Fig.~\ref{fig.amp}.  There are significant
differences between the different curves, which we attribute to intrinsic
variations in the star arising from the stochastic nature of the excitation
and damping.  To investigate this further, we also applied the method to
the combined time series, after first subdividing it into ten 2-day
subsets.  Figure~\ref{fig.amp.segments} shows these amplitude curves and
also their average.

\begin{figure*}
\epsscale{0.8}
\plotone{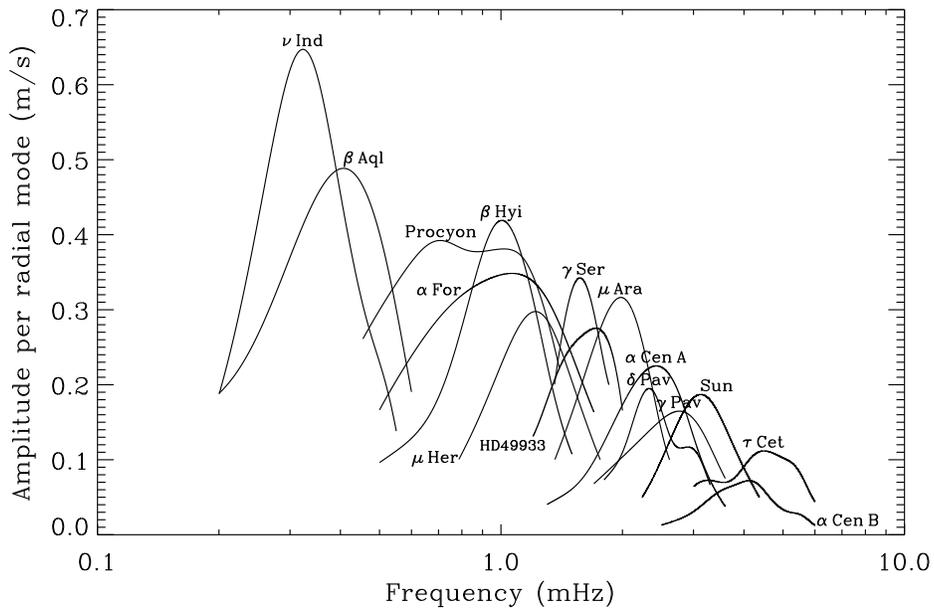}
\caption[]{\label{fig.amp.stars} Smoothed amplitude curves for oscillations
  in Procyon and other stars.
  }
\end{figure*}

The amplitude curve of Procyon has a broad plateau, rather than the single
peak that has been seen for other stars.
\new{Figure~\ref{fig.amp.stars} shows the smoothed amplitude curve for Procyon
compared to the Sun and other stars.  It is an updated version of Fig.~8 of
\citet{KBA2008}, where the following stars have been added: 
$\mu$~Ara \citep{BBS2005}, 
HD~49933 \citep{MBC2005}, 
$\mu$~Her \citep{BBC2008}, 
$\gamma$~Pav \citep{MDM2008}, 
$\tau$~Cet \citep{TKB2008}.
}

The plateau for Procyon is centred at 900\,\muHz{} and is about 500\,\muHz{} wide, with
a mean amplitude across that range of $38.1\pm1.3$\,\cms.  This is our
estimate for the amplitude of the radial ($l=0$) modes in Procyon.
Comparing with the corresponding measurement for the Sun ($18.7\pm
0.7$\,\cms; \citealt{KBA2008}) implies that the velocity oscillations in
Procyon are $2.04 \pm 0.10$ times solar.  In both Procyon and the Sun, the
modes with $l=1$ are higher by a factor of 1.35 (see Table~1 of
\citealt{KBA2008}).

The corresponding intensity amplitude, after accounting for the higher
effective temperature of Procyon (see Eq.~5 in \citealt{K+B95}), is 1.60
times solar.  This implies an amplitude at 500\,nm of 6.8\,ppm for $l=0$
and 8.5\,ppm for $l=1$ (see Table~1 of \citealt{KBA2008}).  These
amplitudes are completely consistent with the detection of oscillations in
Procyon by WIRE \citep{BKB2005b} and with the upper limit set by MOST
\citep{MKG2004,BKB2005}.

The standard deviation of the ten segments in Fig.~\ref{fig.amp.segments}
is $\sigma_A/A = 10.4\% \pm 2.3\%$, which reflects the finite lifetime of
the modes.  We can use equation~(3) from \citet{KBA2008} to
estimate the mode lifetime, but we must account for the much greater width
of the oscillation envelope in Procyon.  
Note that this equation was established empirically and we have confirmed
it analytically using the work of \citet{T+A94}.  We estimate the mode
lifetime to be $\tau = 1.5_{-0.8}^{+1.9}$\,d.  This equals, within rather
large uncertainties, the solar value of 2.9\,d \citep[e.g.,][]{CEI97}.

\section{Conclusions}

We have presented multi-site velocity observations of Procyon that we
obtained with eleven telescopes over more than three weeks.  Combining data
that spans a range of precisions and sampling rates presents a significant
challenge.  When calculating the power spectrum, it is important to use
weights that are based on the measurement uncertainties, otherwise the
result is dominated by the noisiest data.  We have described in detail our
methods for adjusting the weights in order to minimize the noise level in
the final power spectrum.

Our velocity measurements show the clear signature of oscillations.  The
power spectrum shows an excess in a plateau that is centred at 0.9\,mHz and
is broader than has been seen for other solar-type stars.  The mean
amplitude of the radial modes is $38.1\pm1.3$\,\cms{} ($2.04 \pm 0.10$
times solar), which is consistent with previous detections from the ground
and by the WIRE spacecraft, and also with the upper limit set by the MOST
spacecraft.  The variation of the amplitude during the observing campaign
allowed us to estimate the mode lifetime to be $1.5_{-0.8}^{+1.9}$\,d.

We also found a slow variation in the radial velocity of Procyon, with good
agreement between different telescopes.  These variations are remarkably
similar to those seen in the Sun, and we interpret them as being due to
rotational modulation from active regions on the stellar surface.  The variations
appear to have a period of about 10 days, which presumably equals the
stellar rotation period or, perhaps, half of it.  The amount of power in
these slow variations indicates that the fractional area of Procyon
covered by active regions is slightly higher than for the Sun.

The excellent coverage of the observations and the high signal-to-noise
should allow us to produce a good set of oscillation frequencies for
Procyon.  This analysis will be presented in subsequent papers.

\acknowledgments

This work was supported financially by 
the Danish Natural Science Research Council,
the Australian Research Council, 
the Swiss National Science Foundation,
NSF grant AST-9988087 (RPB) and by SUN Microsystems.
We thank Hugh Jones, Chris Tinney and the other members of the
Anglo-Australian Planet Search for agreeing to a time swap that allowed our
AAT observations to be scheduled.
MM is grateful to Prof.\ N. Kameswara Rao, G. Pandey and S. Sriram for
their participation in campaign with VBT Echelle Spectrometer, which was
used for the first time for Doppler spectroscopy observations.

\clearpage

\begin{table*}
\small
\caption{\label{tab.sites} Participating Telescopes}
\begin{tabular}{llllr}
\tableline
\noalign{\smallskip}
\tableline
\noalign{\smallskip}
Identifier & Telescope/Spectrograph & Observatory & Technique & Ref. \\
\noalign{\smallskip}
\tableline
\noalign{\smallskip}
HARPS      & 3.6\,m/HARPS                  & ESO, La Silla,
Chile\footnote{Based on observations collected at the European Southern
Observatory, La Silla, Chile (ESO Programme 078.D-0492 (A)).} 
 & ThAr & 1 \\
CORALIE    & 1.2\,m Euler Telescope/CORALIE  & ESO, La Silla, Chile  & ThAr
& 2 \\
McDonald   & 2.7\,m Harlan J. Smith Tel./coud\'e \'echelle &
                    McDonald Obs., Texas USA  & iodine & 3 \\
Lick       & 0.6\,m CAT/Hamilton \'echelle & Lick Obs., California USA & iodine & 4 \\
UCLES      & 3.9\,m AAT/UCLES & Siding Spring Obs., Australia & iodine & 4 \\
Okayama    & 1.88\,m/HIDES & Okayama Obs., Japan & iodine & 5 \\
Tautenburg & 2\,m/coud\'e \'echelle & Karl Schwarzschild Obs., Germany & iodine & 6 \\
SOPHIE     &  1.93\,m/SOPHIE & Obs.\ de Haute-Provence, France & ThAr & 7 \\
EMILIE     &  1.52\,m/EMILIE+AAA & Obs.\ de Haute-Provence, France & white
light with iodine & 8 \\
SARG       & 3.58\,m TNG/SARG & ORM, La Palma, Spain & iodine & 9 \\
FIES       & 2.5\,m NOT/FIES  & ORM, La Palma, Spain & ThAr & 10 \\
\noalign{\smallskip}
\tableline
\noalign{\smallskip}
\noalign{1.~\citet{RPM2004}; 2.~\citet{B+C2002}; 3.~\citet{ECH2005}; 4.~\cite{BMW96};}
\noalign{5.~\citet{KAS2008}; 6.~\citet{HGK2003}; 7.~\citet{MBM2008};
  8.~\citet{BSB2002} and paper in prep.;  }
\noalign{9.~\citet{CBL2005}; 10.~\citet{F+L2000}}
\end{tabular}
\end{table*}

\begin{table*}
\small
\begin{center}
\caption{\label{tab.obs} Summary of observations}
\begin{tabular}{lrrrrrrrr}
\tableline
\noalign{\smallskip}
\tableline
\noalign{\smallskip}
               & 
Nights     & 
Eff.~Obs.~time & 
 & 
Median $t_{\rm exp}$ & 
Deadtime & 
$f_{\rm Nyq}$ & 
Noise Level & 
Noise per Minute \\
Identifier & 
Allocated  & 
(h)    & 
Spectra &
  (s)  &
  (s) & 
(mHz) & 
(\cms) & 
(\ms)\\
\noalign{
\smallskip}
\tableline
\noalign{
\smallskip}
HARPS      &    8                &  52.0 &  5698 &   5  & 31 & 13.8 &  2.0   & 0.64  \\
CORALIE    &    6                &  27.0 &   936 &  40  & 85 &  4.0 &  9.8   & 2.2   \\
McDonald   &    6                &  16.2 &  1719 &  17  & 15 & 15.6 & 14.4   & 2.5   \\
Lick       &   14                &  95.4 &  1900 & 180  & 54 &  2.1 & 10.9   & 4.7   \\
UCLES      &   12                &  41.4 &  2451 &  16  & 45 &  8.2 &  6.6   & 1.9   \\
Okayama    &   20                &  83.4 &  1997 &  54  &110 &  3.1 &  8.0   & 3.2   \\
Tautenburg &   21                &  14.6 &   494 &  60  & 65 &  4.0 & 22.8   & 3.8   \\
SOPHIE     &    9                &  35.1 &  3924 &  23  & 28 &  9.7 &  4.7   & 1.2   \\
EMILIE     &    4                &  25.7 &  1631 &  47  & 17 &  7.8 & 10.1   & 2.2   \\
SARG       &    4                &  15.2 &   693 &  19  & 65 &  5.9 & 12.6   & 2.1   \\
FIES       &10$\times\frac{1}{2}$&  12.6 &  1087 &  18  & 45 &  7.9 & 21.7   & 3.4   \\
\noalign{                             
\smallskip}                             
\tableline
\end{tabular}
\end{center}
\end{table*}

\clearpage

\end{document}